\documentclass[conference]{IEEEtran}
\IEEEoverridecommandlockouts
\usepackage{cite}
\usepackage{amsmath,amssymb,amsfonts}
\usepackage{algorithm}
\usepackage{algpseudocode}
\usepackage{graphicx}
\usepackage{textcomp}
\usepackage{xcolor}
\usepackage{booktabs} 
\usepackage{multirow} 
\def\BibTeX{{\rm B\kern-.05em{\sc i\kern-.025em b}\kern-.08em
    T\kern-.1667em\lower.7ex\hbox{E}\kern-.125emX}}  
\begin{document}
\title{Explainable Knowledge Distillation for Efficient Medical Image Classification}

\author{\IEEEauthorblockN{1\textsuperscript{st} Aqib Nazir Mir}
\IEEEauthorblockA{\textit{Dept. of Computer Engineering} \\
\textit{Jamia Millia Islamia}\\
New Delhi, India \\
aqib2206142@jmi.ac.in}
\and
\IEEEauthorblockN{2\textsuperscript{nd} Danish Raza Rizvi}
\IEEEauthorblockA{\textit{Dept. of Computer Engineering} \\
\textit{Jamia Millia Islamia}\\
New Delhi, India \\
drizvi@jmi.ac.in}
}

\maketitle

\begin{abstract}
This study comprehensively explores knowledge distillation frameworks for COVID-19 and lung cancer classification using chest X-ray (CXR) images. We employ high-capacity teacher models, including VGG19 and lightweight Vision Transformers (Visformer-S and AutoFormer-V2-T), to guide the training of a compact, hardware-aware student model derived from the OFA-595 supernet. Our approach leverages hybrid supervision, combining ground-truth labels with teacher models’ soft targets to balance accuracy and computational efficiency. We validate our models on two benchmark datasets: COVID-QU-Ex and LCS25000, covering multiple classes, including COVID-19, healthy, non-COVID pneumonia, lung, and colon cancer. To interpret the spatial focus of the models, we employ Score-CAM-based visualizations, which provide insight into the reasoning process of both teacher and student networks. The results demonstrate that the distilled student model maintains high classification performance with significantly reduced parameters and inference time, making it an optimal choice in resource-constrained clinical environments. Our work underscores the importance of combining model efficiency with explainability for practical, trustworthy medical AI solutions.

\end{abstract}


\section{Introduction}
Deep Learning (DL) has revolutionized healthcare by improving diagnostic accuracy and workflow efficiency in modalities such as chest radiography, dermoscopy, histopathology, and cervical cytology \cite{rehman2021survey}. Convolutional Neural Networks (CNNs) and Vision Transformers (ViTs) excel in large-scale image analysis \cite{nissar2024mob,mir2025enhancing}, leveraging datasets like CXR \cite{irvin2019chexpert}, HAM10000 \cite{chang2022ai}, and LCS25000 \cite{borkowski2019lung} for cancer imaging. However, clinical adoption faces two key challenges: the computational complexity of models like DenseNet and ViTs, which hinders deployment in resource-constrained settings, and their “black-box” nature, which reduces trust and interpretability among medical professionals \cite{lin2017focal}.

Knowledge Distillation (KD) tackles the first challenge by transferring knowledge from complex teacher models to lightweight student models, aiming to retain high performance while significantly improving computational efficiency\cite{hinton2015distilling}. In medical imaging, KD has shown promise in applications like chest X-ray (CXR) classification\cite{termritthikun2021eeea} and histopathology\cite{javed2023knowledge}. Techniques include response-based KD, which mimics teacher outputs using Kullback-Leibler divergence\cite{hinton2015distilling}, feature-based KD, which transfers intermediate feature maps\cite{romero2015fitnets,kim2023rckd}, and relation-based KD, which preserves structural relationships\cite{park2019relational}. Recent advancements, such as parameter-efficient KD \cite{rao2023parameter}, federated learning applications\cite{sha2023decoupled}, and hybrid CNN-Transformer models \cite{termritthikun2021eeea}, further enhance KD’s utility.

The second challenge, interpretability, is addressed by Explainable AI (XAI) techniques like CAM\cite{zhou2016learning} and Grad-CAM \cite{selvaraju2017grad}, which visualize model decisions. However, integrating XAI with KD, termed Explainability Distillation (ED), remains underexplored in medical imaging due to high computational costs \cite{sun2025explainability}. Feature-based Distillation (FD) offers a promising, low-cost alternative for enabling explainability, yet its potential as an XAI tool is largely untapped.

This work proposes a novel framework combining feature-based KD with XAI to develop lightweight, interpretable models for medical imaging. By integrating FD with efficient explainability techniques, we aim to enhance clinical trust and enable deployment in resource-limited environments, advancing the practical adoption of DL in healthcare. Our contributions are as follows:

\begin{enumerate}
    \item \textbf{Model Compression via KD:} We train a high-capacity CNN-based teacher model and distill its knowledge into a simplified student model, using Kullback-Leibler (KL) divergence to preserve important feature representations.
    \item \textbf{Cross-Architecture KD:} Unlike traditional approaches where teacher and student share similar structures, we explore heterogeneous architecture transfer to enhance generalization and efficiency.
    \item \textbf{Interpretability Analysis:} We incorporate visualization-based XAI techniques to analyze layer-wise feature maps, by identifying key regions used for classification.
    \item \textbf{Multi-Dataset Validation:} We validate our approach across diverse domains like chest radiography and histopathology datasets, demonstrating its versatility and robustness in handling real-world diagnostic challenges.
\end{enumerate}

Through this work, we aim to answer these questions. Our findings suggest that combining KD with XAI provides a practical pathway to building reliable, resource-efficient AI systems suitable for clinical deployment.

\section{Methodology}

\subsection{Overview of the Proposed Framework}

The proposed explainable knowledge distillation framework is based on the \cite{termritthikun2021eeea}, illustrated in Fig.~1, designed to deploy computationally efficient and interpretable medical image classification models suitable for real-time inference on edge devices. The framework is composed of three primary components:
\begin{itemize}
    \item \textbf{Teacher-Student Architecture:} A robust teacher model with high capacity imparts knowledge to a compact student model via distillation.
    \item \textbf{Distillation Loss Mechanism:} A hybrid loss function combining hard labels and soft teacher outputs supervises the student.
    \item \textbf{Explainability Module:} Visual interpretation methods such as Score-CAM provide clinical transparency and trust.
\end{itemize}
This architecture is motivated by the challenge of deploying traditional deep models (e.g., CNNs and Transformers), which are often computationally expensive, on resource-constrained devices. Knowledge distillation enables a smaller model to acquire diagnostic capabilities from a larger, high-performing model while preserving strong performance.

\begin{figure*}[t]
\centering
\includegraphics[width=\textwidth]{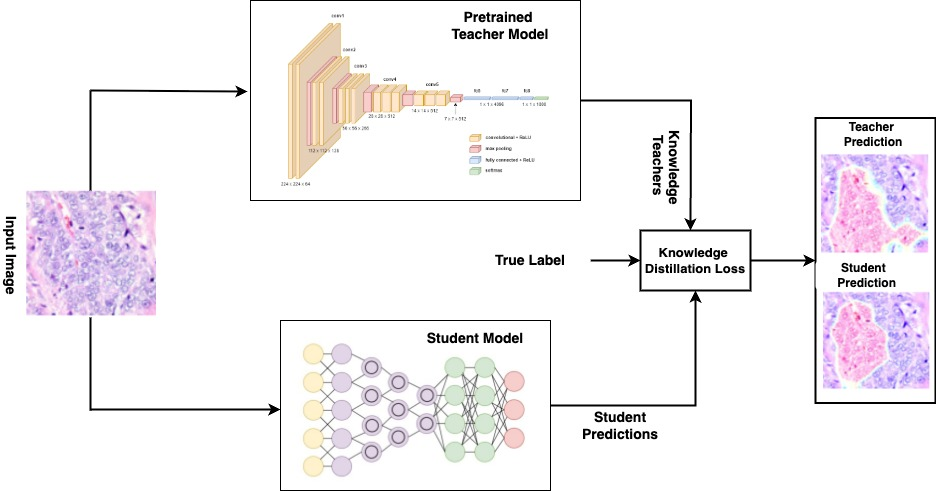}
\caption{Architecture of the proposed model. Knowledge is transferred from a selected teacher model to a lightweight student, while Score-CAM provides visual explanations. One teacher is chosen at a time from the pool of teachers.}
\label{fig:2nd}
\end{figure*}

\subsection{Teacher and Student Model Architectures}

\subsubsection{Teacher Models}

To investigate the effectiveness of knowledge transfer across architectural paradigms, we employ three diverse and representative teacher networks:
\begin{itemize}
    \item \textbf{VGG19:} A deep convolutional neural network with 16 convolutional layers arranged into sequential blocks, each comprising 3$\times$3 convolution filters, ReLU activations, and max pooling operations. Batch normalization layers stabilize training, followed by a global average pooling layer. Pretrained ImageNet weights initialize convolutional layers, while the classification head is trained from scratch on the target dataset.
    \item \textbf{Visformer-S:} A hybrid architecture that integrates convolutional inductive bias into vision transformers by combining early convolutional layers with transformer blocks. This design enhances spatial locality while preserving the transformer’s ability to model long-range dependencies. Visformer-S is lightweight and performs well in medical imaging, making it suitable for teaching.
    \item \textbf{AutoFormer-V2-T:} A highly efficient transformer-based model designed via neural architecture search. AutoFormer-V2-T is tailored for resource-efficient deployment and is a robust baseline for studying the transfer of attention-based features during distillation.
\end{itemize}

These teacher networks cover a wide architectural spectrum from purely convolutional (VGG19) to attention-centric (AutoFormer-V2-T), providing insights into how different representational strategies influence student learning.

\subsubsection{Student Model}

\textbf{OFA-595 (Once-For-All Subnet):}  
The student network is a compact, hardware-aware architecture derived from the OFA supernet using progressive shrinking. The OFA-595 subnet is optimized for Pareto efficiency, balancing inference latency and accuracy, reflecting realistic deployment on edge devices. It incorporates depthwise separable convolutions and elastic kernel sizes, enabling adaptive capacity and low computational overhead. All student weights are initialized from scratch to ensure unbiased knowledge transfer via distillation.

\subsection{Training Procedure}

The training pipeline has two sequential phases:
\begin{itemize}
    \item \textbf{Phase 1: Teacher Pretraining} \\
    Each teacher network is trained independently on domain-specific datasets (e.g., histopathological images and chest X-rays) using Binary Cross-Entropy (BCE) loss. To prevent overfitting, we used regularization and dropout techniques. This phase involves training on large datasets such as LCS25000 and COVID-QU to ensure robust feature learning and generalizability. This results in well-trained teacher models that provide informative soft targets to guide the student models.
    
    \item \textbf{Phase 2: Knowledge Distillation } \\
    The student model (OFA-595) is trained under the guidance of pretrained teacher models using a hybrid loss function, which integrates the binary cross-entropy loss with a distillation loss that aligns the student’s predictions with the softened logits of the teacher.
    \begin{equation}
    L_{KD} = \alpha \cdot L_{CE}(y, \hat{y}_s) + (1-\alpha) \cdot L_{KL}(\mathrm{soft}(y_t), \mathrm{soft}(\hat{y}_s)),
    \end{equation}
    where $\hat{y}_s$ and $y_t$ are student and teacher logits, and $\alpha$ balances supervision components. Temperature scaling softens teacher outputs for richer gradients. All student results (performance, latency, explainability) use this distillation setup.
\end{itemize}

\begin{algorithm}
\caption{Knowledge Distillation Procedure for OFA-595(Student Model)}
\label{alg:knowledge_distillation}
\begin{algorithmic}[1]
\State \textbf{Input:}
\State \quad Training dataset $\mathcal{D} = \{(x_i, y_i)\}_{i=1}^N$
\State \quad Pre-trained teacher model $f_t$ (e.g., DenseNet161)
\State \quad Student model $f_s$ (e.g., OFA-595)
\State \quad Balancing coefficient $\alpha \in [0, 1]$
\State \textbf{Output:}
\State \quad Trained student model $f_s$
\State \textbf{Procedure:}
\For{each mini-batch $(x, y) \in \mathcal{D}$}
    \State \textbf{Forward Pass:}
    \State \quad Compute teacher output logits: $p_t = f_t(x)$
    \State \quad Compute student output logits: $p_s = f_s(x)$
    \State \textbf{Loss Computation:}
    \State \quad Compute Focal Binary Cross-Entropy loss:
    \State \quad \quad $\mathcal{L}_{\text{FBCE}} = \text{FBCE}(p_s, y)$
    \State \quad Compute Mean Squared Error loss:
    \State \quad \quad $\mathcal{L}_{\text{MSE}} = \text{MSE}(p_s, p_t)$
    \State \textbf{Combined Knowledge Distillation Loss:}
    \State \quad $\mathcal{L}_{\text{KD}} = \alpha \cdot \mathcal{L}_{\text{FBCE}} + (1 - \alpha) \cdot \mathcal{L}_{\text{MSE}}$
    \State \textbf{Optimization Step:}
    \State \quad Backpropagate the loss $\mathcal{L}_{\text{KD}}$
    \State \quad Update the student model $f_s$
\EndFor
\end{algorithmic}
\end{algorithm}

\subsection{Loss Function for Knowledge Distillation}
\label{subsec:loss_function}
In this work, we adopt a knowledge distillation framework designed for multi-label classification tasks, where a lightweight student network learns from both hard labels and the soft predictions of a larger, pre-trained teacher network. The key objective is to facilitate knowledge transfer by enabling the student model to replicate the teacher’s behavior while learning effectively from the ground truth annotations.

For multi-class problems, the output layer of a neural network typically employs a sigmoid activation function to generate class-wise probability scores ranging between 0 and 1. The predicted probabilities and binary ground truth labels are utilized to calculate the BCE loss, the primary supervised loss function. The BCE loss for a single instance can be evaluated as:
\begin{equation}
L_{BCE}(p, y) = \begin{cases} 
-\log(p), & \text{if } y = 1 \\ 
-\log(1 - p), & \text{if } y = 0 
\end{cases}
\label{eq:bce_loss}
\end{equation}
where \( p \) represents the predicted probability for a given class and \( y \) denotes the corresponding ground truth label. When focusing on the student model prediction \( p_s \), the BCE loss simplifies to:
\begin{equation}
L_{BCE}(p_s) = -\log(p_s),
\label{eq:bce_simplified}
\end{equation}
depending on whether the ground truth label is 0 or 1.

A major challenge in multi-label medical image classification is the issue of class imbalance, where certain pathological categories may be underrepresented. To address this, we integrate a Focal Binary Cross-Entropy (FBCE) loss, which extends the standard BCE loss by incorporating a modulating factor to focus learning on harder, misclassified examples. The FBCE loss is defined as:
\begin{equation}
L_{FBCE}(p_s, y) = \begin{cases} 
-(1 - p_s)^\gamma \log(p_s), & \text{if } y = 1 \\ 
-(p_s)^\gamma \log(1 - p_s), & \text{if } y = 0 
\end{cases}
\label{eq:fbce_loss}
\end{equation}
Here, \( \gamma \) is a focusing parameter that controls the degree of emphasis on difficult samples. Higher values of \( \gamma \) increase the down-weighting effect on well-classified examples, encouraging the model to pay greater attention to challenging cases. When \( \gamma = 0 \), the FBCE loss becomes equivalent to the standard BCE loss.

To enable the student network to mimic the soft output distribution of the teacher, we utilize Mean Squared Error (MSE) loss for distillation. This soft loss encourages the student to align its prediction scores with the teacher’s, which contain richer information than binary labels alone. Before computing MSE, both student and teacher logits are passed through a sigmoid function, scaled by a temperature factor \( T \), to produce softened probabilities:
\begin{equation}
L_{MSE}(p_s, p_t) = \frac{T^2}{n} \sum_{i=1}^{n} \left( \sigma\left(\frac{p_{s,i}}{T}\right) - \sigma\left(\frac{p_{t,i}}{T}\right) \right)^2
\label{eq:mse_loss}
\end{equation}
where \( p_s \) and \( p_t \) are the student and teacher logits, respectively, \( n \) is the number of labels, and \( \sigma(\cdot) \) denotes the sigmoid activation function. The temperature \( T \) regulates the softness of the output distribution, enabling more effective knowledge transfer.

The final loss function used for training the student model is a linear combination of the FBCE loss (reflecting supervision from hard labels) and the MSE loss (capturing soft supervision from the teacher). This combined loss is given by:
\begin{equation}
L_{KD} = \alpha \cdot L_{FBCE}(p_s, y) + (1 - \alpha) \cdot L_{MSE}(p_s, p_t)
\label{eq:kd_loss}
\end{equation}
where \( \alpha \in [0,1] \) is a balancing coefficient that controls the relative importance of hard and soft targets. In our experiments, we set \( \alpha = 0.5 \), following standard practice, to ensure an equal contribution from both types of supervision.

This hybrid loss function enables the student network to learn not only from the ground truth annotations but also from the nuanced predictions of the teacher, resulting in improved generalization and performance, particularly in scenarios with limited or imbalanced data.

\section{Experimental Results}
\subsection{Datasets}
 We employ two accessible and carefully curated medical imaging datasets: COVID-QU-Ex\cite{tahir2022covid}and LCS25000\cite{borkowski2019lung}. The COVID-QU dataset [12] contains 33,920 CXR images, with 11,956 confirmed cases, including viral or bacterial pneumonia. Per the original authors’ recommendation, we adopt a 65/15/20 split for training, validation, and testing for the COVID-QU dataset. The COVID dataset includes 10,701 healthy images and 11,263 images of non-COVID infections, such as pneumonia, with a 65/15/20 split for training, validation, and testing. Additionally, the LCS25000 dataset comprises 25,000 histopathological image patches, evenly divided into 12,500 benign and 12,500 malignant classes. These exhibit variability in magnification and staining, making the dataset suitable for robust classification tasks. For this dataset, we adopt a 70/10/20 split for training, validation, and test subsets in alignment with common practice.

\subsection{Baseline Evaluation on Benchmark Datasets and Implementation}
To evaluate the efficacy of the proposed OFA-595 architecture, we conducted a comparative analysis against several widely used convolutional and transformer-based models, namely VGG-19, DenseNet-201, EEEA-Net-C2, AutoFormerV2, and Visformer-S. The evaluation was performed on two publicly available datasets: the LCS25000 dataset, which includes lung and colon cancer cases, and the COVID-QU-Ex dataset, comprising COVID, non-COVID, and pneumonia cases.

The teacher and student models were developed using the PyTorch deep learning framework, employing standard implementations with default hyperparameters unless stated otherwise. The models were trained using the Adam optimizer, which is well-suited for handling weight decay in transformer-based architectures. Input images were uniformly resized to a resolution of 224×224 pixels to maintain compatibility with the backbone networks. A batch size of 32 was used to ensure efficient and stable gradient updates. The training pipeline was divided into two distinct phases: initially, the teacher model was trained independently for 10 epochs to convergence. In the subsequent phase, the pre-trained teacher model was employed to supervise the student model using a knowledge distillation framework, and the student was trained for an additional 10 epochs. 

\begin{table}[t]
\caption{Performance comparison of models on LCS25000 and COVID-QU-Ex datasets.}
\label{tab:baseline_performance}
\centering
\begin{tabular}{l|ccc|ccc}
\toprule
 & \multicolumn{3}{c|}{\textbf{LCS25000}} & \multicolumn{3}{c}{\textbf{COVID-QU-Ex}} \\
\cmidrule(lr){2-4} \cmidrule(lr){5-7}
\textbf{Model} & \textbf{AUC} & \textbf{Acc} & \textbf{F1} & \textbf{AUC} & \textbf{Acc} & \textbf{F1} \\
\midrule
VGG-19 & 81.1 & 91.2 & 88.3 & 88.1 & 94.3 & 90.1 \\
DenseNet-201 & 84.1 & 94.2 & 94.1 & 89.1 & 96.8 & 96.3 \\
OFA-595 & 83.1 & 96.1 & 94.3 & 90.2 & 96.0 & 96.4 \\
Visformer-S & 82.4 & 96.5 & 94.7 & 91.4 & 97.3 & 97.1 \\
EEENA-Net C2 & 82.9 & 95.9 & 94.6 & 89.9 & 96.8 & 96.9 \\
\bottomrule
\end{tabular}
\end{table}

Table~\ref{tab:baseline_performance} summarizes the results of each model in terms of AUC, Accuracy, and F1-score.On the LCS25000 dataset, Visformer-S demonstrated the highest classification accuracy (96.5\%)and F1-score (94.7\%), which can be attributed to its efficient hybrid vision-transformer architecture that captures local and global dependencies. Interestingly, DenseNet-201 achieved the highest AUC (84.1\%), reflecting its strong confidence in separating class distributions. The proposed EEEA-Net-C2, despite having a more compact and efficient structure, achieved a highly competitive performance (AUC: 82.9\%, Acc: 95.9\%, F1: 94.6\%), closely rivaling these heavier models while using fewer parameters. This highlights the effectiveness of the EEEA block in enabling rich feature extraction while maintaining computational efficiency.

On the COVID-QU dataset, the superiority of transformer-based architectures became more evident. Visformer-S again led with the highest AUC (91.4\%), accuracy (97.3\%), and a strong F1-score (97.1\%). The EEEA-Net-C2 model, however,  demonstrated remarkable generalization, attaining an AUC of 89.9\%, accuracy of 96.8\%, and F1-score of 96.9\%. This confirms that EEEA-Net-C2 can achieve high diagnostic accuracy across data sets of varying quality and complexity, particularly when optimized.

\subsection{Effectiveness of Knowledge Distillation}
We implemented a knowledge distillation strategy to optimize model performance while maintaining a lightweight footprint. In this setup, various high-capacity models (e.g., DenseNet-201, AutoFormerV2, Visformer-S) served as teacher networks, transferring learned representations to a more compact student model (primarily OFA-595, which shares architectural similarity with EEEA-Net-C2).
\begin{table}[t]
\centering
\caption{Performance of OFA-595 student with various teachers}
\label{tab:kd_performance}
\begin{tabular}{l ccc ccc}
\toprule
 & \multicolumn{3}{c}{\textbf{LCS25000 Dataset}} & \multicolumn{3}{c}{\textbf{COVID-QU-Ex Dataset}} \\
\cmidrule(lr){2-4} \cmidrule(lr){5-7}
\textbf{Teacher} & \textbf{AUC} & \textbf{Acc.} & \textbf{F1} & \textbf{AUC} & \textbf{Acc.} & \textbf{F1} \\
\midrule
VGG-19 & 82.9 & 92.4 & 93.1 & 86.1 & 94.3 & 94.1 \\
DenseNet-201 & 85.1 & 93.8 & 95.6 & 88.9 & 96.8 & 96.2 \\
OFA-595 & 85.9 & 95.2 & 96.0 & 91.1 & 96.7 & 97.3 \\
Visformer-S & 83.6 & 97.2 & 95.09 & 92.0 & 97.0 & 97.1 \\
EEENA-Net C2 & 83.2 & 96.1 & 96.9 & 93.7 & 96.9 & 98.7 \\
AutoFormer & 83.7 & 95.6 & 95.4 & 89.9 & 97.0 & 97.5 \\
\bottomrule
\end{tabular}
\end{table}

As shown in Table~\ref{tab:kd_performance}, the DenseNet-201 teacher yielded consistent improvement in student performance across both datasets. On the LCS25000 dataset, using DenseNet-201 as the teacher led to an improvement in AUC (85.1\%), accuracy (93.8\%), and F1-score (95.6\%) for the OFA-595 student, indicating effective transfer of both feature representations and decision boundaries. On the COVID-QU-Ex dataset, the same teacher-student configuration further improved performance to 88.9\% AUC, (96.8\%) accuracy, and 96.2\% F1-score.

Notably, AutoFormer and Visformer-S also contributed positively as teachers. The AutoFormer–OFA-595 configuration achieved an F1-score of 97.5\% and accuracy of 97.0\% on the COVID-QU-Ex dataset, highlighting the potential of transformer-based teachers in transferring rich contextual information. On the other hand, Visformer-S delivered the highest accuracy (97.2\%) on the LCS25000 dataset with an F1-score of 95.09\%, demonstrating its strength in structural representation learning.

Furthermore, the EEEA-Net-C2 model itself, when used as a teacher, guided the OFA-595 student to impressive results achieving 93.7\% AUC, 96.9\% accuracy, and the highest F1-score of 98.7\% on the COVID-QU-Ex dataset. This underscores the model’s robustness and applicability even in teacher roles, affirming its value beyond standalone use. These findings highlight that the effectiveness of knowledge distillation is highly dependent on the choice of the teacher. 

\begin{figure}[t]
\centering
\includegraphics[width=\columnwidth]{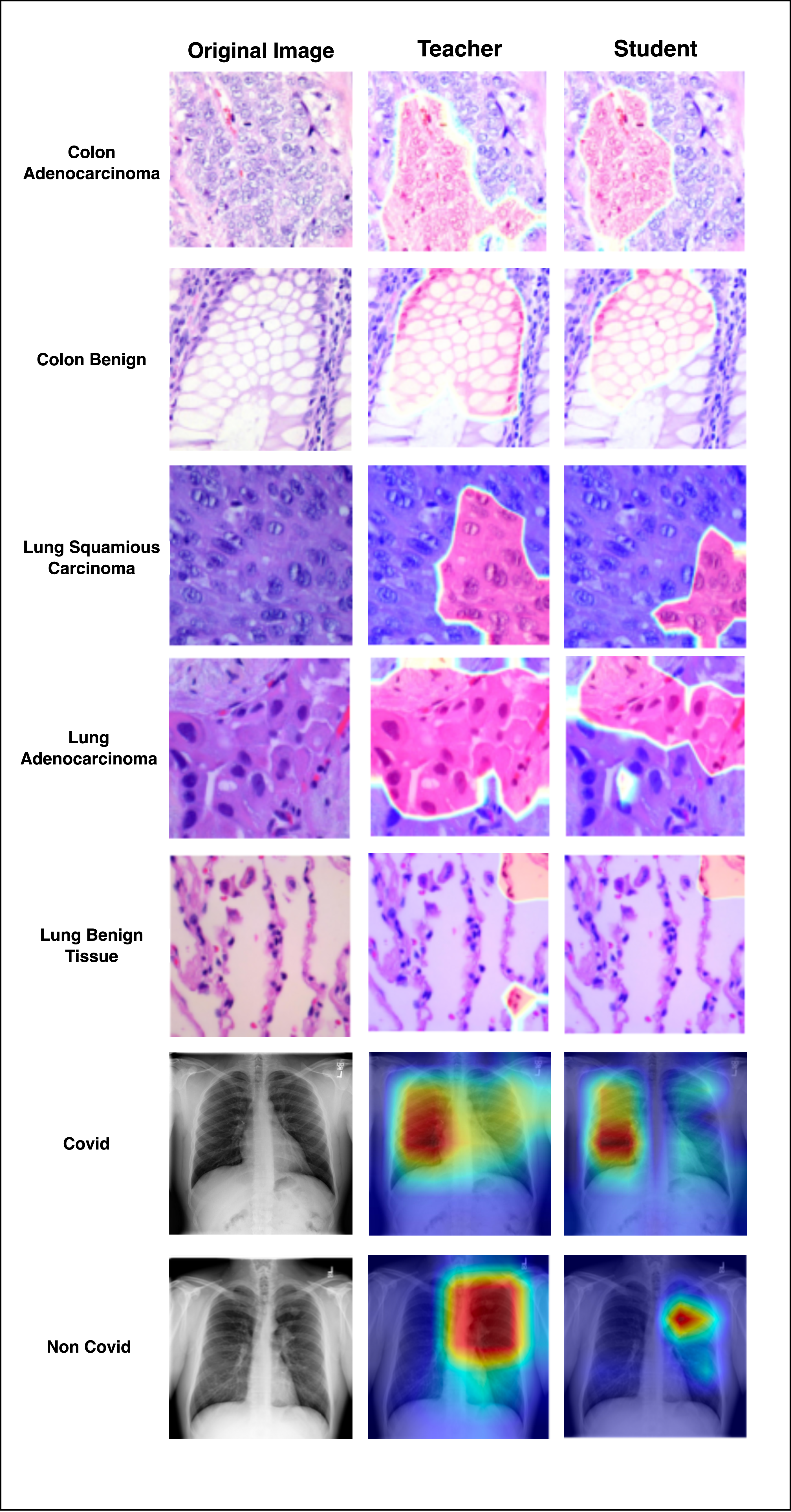}
\caption{Score-CAM heatmaps for teacher and student networks across histopathological and radiological images.}
\label{fig:score_cam}
\end{figure}

\subsection{Explainable AI for Knowledge Distillation}
We visualize the heatmaps generated using Score-CAM for histopathological and radiological images to interpret knowledge transfer from teacher to student networks. These visualizations allow us to qualitatively assess how well the student model replicates the spatial attention of the teacher model across different diagnostic categories.

As illustrated in Fig.~\ref{fig:score_cam}, the first column shows the original input image, while the second and third columns represent the corresponding heatmaps produced by the teacher and student networks, respectively. The figure includes examples from a variety of disease classes.

Teachers and student models consistently focus on diagnostically relevant regions for the histopathological images. For instance, in Colon Adenocarcinoma and Lung Squamous Carcinoma, the highlighted heatmap regions align well with dense cell structures and abnormal tissue patterns. The student model successfully mimics the teacher’s attention by focusing on similar pathological zones, validating the effectiveness of knowledge distillation in capturing critical features.

In the radiological domain, particularly in COVID and non-COVID chest X-ray cases, the Score-CAM maps show that the teacher network attends to broader lung regions associated with infection indicators. Meanwhile, the student model’s attention becomes more refined and localized, particularly in the non-COVID case, where it highlights regions that better correspond to subtle anatomical changes.

Overall, the consistent alignment of heatmaps between the student and teacher networks demonstrates the ability of the student model to internalize the discriminative spatial features guided by the teacher. These visual results confirm that knowledge distillation improves quantitative performance and effectively transfers interpretable, clinically meaningful attention patterns from teacher to student networks.
\subsection{Limitations}
Our framework, while effective on COVID-QU-Ex and LCS25000 datasets, may overfit to their specific characteristics, such as patient demographics or imaging protocols, potentially limiting generalizability to unseen datasets like CheXpert or ISIC 2019 with different disease distributions. The student model’s reliance on teacher models (e.g., DenseNet-201, Visformer-S) risks propagating biases or overfitting from the teachers, reducing robustness. Additionally, deploying the OFA-595 model in resource-constrained settings may require further optimization for low-end edge devices. Finally, while Score-CAM enhances interpretability, it may not fully capture complex clinical features, potentially affecting trust in high-precision scenarios. Future work will validate the framework on diverse datasets and optimize for edge deployment.
\section{Conclusion}

This study presents an effective knowledge distillation framework integrating model compression with explainable AI to enable efficient and interpretable medical image classification. By distilling knowledge from high-capacity teacher models, such as DenseNet-201, Visformer-S, and AutoFormer-V2-T, into a lightweight OFA-595 student model, we achieved robust performance on two medical imaging datasets: COVID-QU-Ex for chest X-rays and LCS25000 for histopathology. The student model demonstrated competitive metrics, with accuracies up to 97.0\% and F1-scores up to 97.5\% on COVID-QU-Ex, closely rivaling heavier models while significantly reducing computational demands. To enhance interpretability, Score-CAM visualizations showed that the student model successfully replicated the teacher’s attention to clinically relevant regions, such as infection sites and dense cell structures, supporting transparency and clinical trust. Our contributions include cross-architecture KD for improved generalization, a hybrid loss function addressing class imbalance, and validation on diverse medical imaging modalities, demonstrating the framework’s versatility.

Future work will focus on extending our knowledge distillation framework for clinical adoption. We plan to validate its performance on diverse medical imaging datasets and conduct pilot studies in hospital settings to assess diagnostic accuracy and workflow efficiency for conditions such as tuberculosis and pneumonia. To enhance interpretability, we will integrate advanced XAI techniques like Grad-CAM++ for more precise visualization of clinically relevant features. The student model will be optimized for low-power edge devices, targeting real-time diagnostics with sub-100ms latency in resource-constrained environments. Additionally, we will align the framework with regulatory standards (e.g., FDA) and integrate it into hospital systems such as PACS, ensuring seamless deployment and scalability across diverse healthcare infrastructures.
\bibliographystyle{IEEEtran}
\bibliography{references}

\begin{thebibliography}{10}
\providecommand{\url}[1]{#1}
\csname url@samestyle\endcsname
\providecommand{\newblock}{\relax}
\providecommand{\bibinfo}[2]{#2}
\providecommand{\BIBentrySTDinterwordspacing}{\spaceskip=0pt\relax}
\providecommand{\BIBentryALTinterwordstretchfactor}{4}
\providecommand{\BIBentryALTinterwordspacing}{\spaceskip=\fontdimen2\font plus
\BIBentryALTinterwordstretchfactor\fontdimen3\font minus \fontdimen4\font\relax}
\providecommand{\BIBforeignlanguage}[2]{{%
\expandafter\ifx\csname l@#1\endcsname\relax
\typeout{** WARNING: IEEEtran.bst: No hyphenation pattern has been}%
\typeout{** loaded for the language `#1'. Using the pattern for}%
\typeout{** the default language instead.}%
\else
\language=\csname l@#1\endcsname
\fi
#2}}
\providecommand{\BIBdecl}{\relax}
\BIBdecl

\bibitem{rehman2021survey}
A.~Rehman, M.~A. Butt, and M.~Zaman, ``A survey of medical image analysis using deep learning approaches,'' in \emph{2021 5th International Conference on Computing Methodologies and Communication (ICCMC)}.\hskip 1em plus 0.5em minus 0.4em\relax IEEE, 2021, pp. 1334--1342.

\bibitem{nissar2024mob}
I.~Nissar, S.~Alam, S.~Masood, and M.~Kashif, ``Mob-cbam: A dual-channel attention-based deep learning generalizable model for breast cancer molecular subtypes prediction using mammograms,'' \emph{Computer Methods and Programs in Biomedicine}, vol. 248, p. 108121, 2024.

\bibitem{mir2025enhancing}
A.~N. Mir, D.~R. Rizvi, and M.~R. Ahmad, ``Enhancing histopathological image analysis: An explainable vision transformer approach with comprehensive interpretation methods and evaluation of explanation quality,'' \emph{Engineering Applications of Artificial Intelligence}, vol. 149, p. 110519, 2025.

\bibitem{irvin2019chexpert}
J.~Irvin, P.~Rajpurkar, M.~Ko, Y.~Yu, S.~Ciurea-Ilcus, C.~Chute, H.~Marklund, B.~Haghgoo, R.~Ball, K.~Shpanskaya \emph{et~al.}, ``Chexpert: A large chest radiograph dataset with uncertainty labels and expert comparison,'' in \emph{Proceedings of the AAAI Conference on Artificial Intelligence}, vol.~33, no.~01, 2019, pp. 590--597.

\bibitem{chang2022ai}
C.~H. Chang, W.~E. Wang, F.~Y. Hsu, R.~J. Chen, and H.~C. Chang, ``Ai ham 10000 database to assist residents in learning differential diagnosis of skin cancer,'' in \emph{2022 IEEE 5th Eurasian Conference on Educational Innovation (ECEI)}.\hskip 1em plus 0.5em minus 0.4em\relax IEEE, 2022, pp. 1--3.

\bibitem{borkowski2019lung}
A.~A. Borkowski, M.~M. Bui, L.~B. Thomas, C.~P. Wilson, L.~A. DeLand, and S.~M. Mastorides, ``Lung and colon cancer histopathological image dataset (lc25000),'' \emph{arXiv preprint arXiv:1912.12142}, 2019.

\bibitem{lin2017focal}
T.-Y. Lin, P.~Goyal, R.~Girshick, K.~He, and P.~Dollár, ``Focal loss for dense object detection,'' in \emph{Proceedings of the IEEE International Conference on Computer Vision}, 2017, pp. 2980--2988.

\bibitem{hinton2015distilling}
G.~Hinton, O.~Vinyals, and J.~Dean, ``Distilling the knowledge in a neural network,'' \emph{arXiv preprint arXiv:1503.02531}, 2015.

\bibitem{termritthikun2021eeea}
C.~Termritthikun, Y.~Jamtsho, J.~Ieamsaard, P.~Muneesawang, and I.~Lee, ``Eeea-net: An early exit evolutionary neural architecture search,'' \emph{Engineering Applications of Artificial Intelligence}, vol. 104, p. 104397, 2021.

\bibitem{javed2023knowledge}
S.~Javed, A.~Mahmood, T.~Qaiser, and N.~Werghi, ``Knowledge distillation in histology landscape by multi-layer features supervision,'' \emph{IEEE Journal of Biomedical and Health Informatics}, vol.~27, no.~4, pp. 2037--2046, 2023.

\bibitem{romero2015fitnets}
A.~Romero, N.~Ballas, S.~Kahou, A.~Chassang, C.~Gatta, and Y.~Bengio, ``Fitnets: Hints for thin deep nets,'' 2015.

\bibitem{kim2023rckd}
H.~Kim, T.~Kwak, H.~Chang, S.~Kim, and I.~Kim, ``Rckd: Response-based cross-task knowledge distillation for pathological image analysis,'' \emph{Bioengineering}, vol.~10, no.~11, p. 1279, 2023.

\bibitem{park2019relational}
W.~Park, D.~Kim, Y.~Lu, and M.~Cho, ``Relational knowledge distillation,'' 2019.

\bibitem{rao2023parameter}
J.~Rao, X.~Meng, L.~Ding, S.~Qi, X.~Liu, M.~Zhang, and D.~Tao, ``Parameter-efficient and student-friendly knowledge distillation,'' \emph{IEEE Transactions on Multimedia}, 2023.

\bibitem{sha2023decoupled}
X.~Sha, Y.~Wang, and T.~Fang, ``Decoupled knowledge distillation in data-free federated learning,'' in \emph{International Artificial Intelligence Conference}.\hskip 1em plus 0.5em minus 0.4em\relax Springer, 2023, pp. 164--177.

\bibitem{zhou2016learning}
B.~Zhou, A.~Khosla, A.~Lapedriza, A.~Oliva, and A.~Torralba, ``Learning deep features for discriminative localization,'' in \emph{Proceedings of the IEEE Conference on Computer Vision and Pattern Recognition}, 2016, pp. 2921--2929.

\bibitem{selvaraju2017grad}
R.~Selvaraju, A.~Das, R.~Vedantam, M.~Cogswell, D.~Parikh, and D.~Batra, ``Grad-cam: Why did you say that?'' 2017.

\bibitem{sun2025explainability}
T.~Sun, H.~Chen, G.~Hu, and C.~Zhao, ``Explainability-based knowledge distillation,'' \emph{Pattern Recognition}, vol. 159, p. 111095, 2025.

\bibitem{tahir2022covid}
A.~M. Tahir, M.~E. Chowdhury, Y.~Qiblawey, A.~Khandakar, T.~Rahman, S.~Kiranyaz, U.~Khurshid, N.~Ibtehaz, S.~Mahmud, and M.~Ezeddin, ``Covid-qu-ex dataset,'' \emph{Kaggle}, 2022.

\end{thebibliography}

\end{document}